\documentclass[letterpaper]{article} %
\usepackage{ADReasoner} 
\usepackage[hyphens]{url}  %
\usepackage{graphicx} %
\usepackage{natbib}  %
\usepackage{caption} %
\usepackage{algorithm}
\usepackage{algorithmic}
\usepackage{amsmath}
\usepackage{amssymb}
\usepackage{colortbl}
\usepackage{newfloat}
\usepackage{listings}
\DeclareCaptionStyle{ruled}{labelfont=normalfont,labelsep=colon,strut=off} %
\floatstyle{ruled}
\newfloat{listing}{tb}{lst}{}
\floatname{listing}{Listing}

\usepackage{booktabs}

\usepackage{multirow}
\usepackage{makecell}
\usepackage{array}
\usepackage{threeparttable}
\usepackage{pifont}

\newcommand{\cmark}{\ding{51}}
\newcommand{\xmark}{\ding{55}}

\title{ADReasoner: Acoustic-Linguistic Cue-Aware Reinforcement Learning for Explainable Alzheimer's Disease Detection from Speech}

\author{
    Yin-Long Liu\textsuperscript{\rm 1},
    Yuanchao Li\textsuperscript{\rm 2},
    Jiahong Yuan\textsuperscript{\rm 1},
    Zhen-Hua Ling\textsuperscript{\rm 1}
}

\affiliations{
    \textsuperscript{\rm 1}University of Science and Technology of China, Hefei, Anhui, China\\
    \textsuperscript{\rm 2}University of Cambridge, Cambridge, England, UK\\
    lyl2001@mail.ustc.edu.cn
}

\begin{document}

\maketitle

\begin{abstract}
Speech-based Alzheimer’s Disease (AD) detection holds significant promise due to its scalable and non-invasive nature. However, current Speech Large Language Models (SpeechLLMs), similar to conventional AD detection systems, still treat AD detection primarily as a simple categorical classification task. This paradigm provides limited interpretability of predictions, while leaving the inherent expressiveness and reasoning capabilities of SpeechLLMs largely underutilized. In this work, we take the first step toward reframing AD detection as a deep reasoning task via Reinforcement Learning (RL). We propose ADReasoner, an RL-enhanced SpeechLLM designed to generate accurate diagnostic predictions with interpretable rationales grounded in fine-grained acoustic-linguistic cues. To achieve this, we first systematically construct ADCoT-7K, an AD speech reasoning dataset combining real data and our quality-filtered synthetic data, where each instance features Chain-of-Thought (CoT) annotations and detailed captions. Second, observing the weak perception of AD-related acoustic-linguistic cues in current SpeechLLMs, we leverage Supervised Fine-Tuning (SFT) to develop a cue-perception-enhanced foundation model, ADReasoner-Base, thereby improving AD speech understanding. Third, alongside two standard rule-based rewards, we introduce a reasoning reward and a reasoning-label consistency reward into Group Relative Policy Optimization (GRPO) for RL. Furthermore, a shrinkage mechanism is employed to suppress misleading reasoning signals, while a gating mechanism progressively introduces both rewards to stabilize policy training. Experiments on three AD datasets (Pitt, PROCESS-1, PROCESS-2) demonstrate that ADReasoner outperforms previous state-of-the-art models in both detection performance and explanation quality, advancing AD detection toward interpretable reasoning from speech.
\end{abstract}

\begin{figure}[t]
    \centering
    \includegraphics[width=1\linewidth]{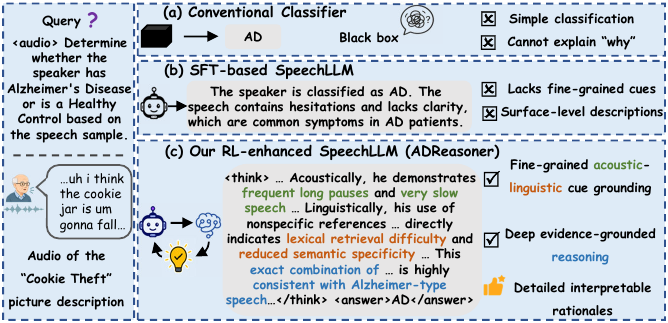}
    \caption{In contrast to (a) conventional classifiers that merely output discrete labels and (b) SFT-based approaches confined to surface-level descriptions, (c) our ADReasoner generates diagnostic predictions with detailed interpretable rationales grounded in fine-grained acoustic-linguistic cues.}
    \label{fig:classifier_vs_sft_vs_rl}
\end{figure}

\section{Introduction}

Alzheimer’s Disease (AD), the leading cause of dementia, causes progressive decline in cognitive functions such as memory and language \cite{nestor2004advances}.
Due to its subtle onset, early detection is critical for timely treatment, yet traditional diagnostics such as lumbar puncture are often invasive, costly, and resource-intensive \cite{de2019brief,kourtis2019digital}.
Since speech and language impairments are prominent early signs of AD, their analysis has emerged as a non-invasive and scalable detection approach \cite{beltrami2018speech}. 
Consequently, extensive research has explored acoustic and linguistic biomarkers for distinguishing individuals with AD from Healthy Controls (HC) \cite{rohanian21_interspeech,tao2025early}.
Nevertheless, conventional speech-based AD detection models typically operate as black boxes, merely outputting discrete binary labels (AD vs. HC) without explaining the underlying evidence \cite{shankar2025systematic}.

With the rapid advancement of Speech Large Language Models (SpeechLLMs) across downstream tasks \cite{peng2025survey}, researchers have begun applying them to AD detection.
However, most existing studies still formulate AD detection as a categorical classification task \cite{shahin2025zero,liu2026cross}. This limits interpretability and underutilizes the rich expressiveness and reasoning capabilities of SpeechLLMs. 
Although a few works \cite{feng2025cogbench,feng2026cross} have explored preliminary explainability by requiring SpeechLLMs to generate diagnostic labels alongside rationales, these rationales are learned solely via Supervised Fine-Tuning (SFT), lacking direct optimization through task-specific Reinforcement Learning (RL). 
Consequently, they may remain largely confined to surface-level descriptions \cite{wang2026emotionthinker}, without systematically connecting fine-grained acoustic–linguistic cues to their clinical significance. As a result, current SpeechLLMs still struggle to generate in-depth, cue-grounded diagnostic reasoning.

In this work, we take the first step toward RL-based explainable AD detection. To this end, we identify four critical challenges: (i) \textit{Severe data scarcity}: public AD speech datasets are highly limited in size due to privacy constraints and collection costs. (ii) \textit{Lack of reasoning annotations}: existing datasets lack fine-grained acoustic-linguistic captions and explicit Chain-of-Thought (CoT) annotations for structured reasoning. (iii) \textit{Weak cue perceptions}: current SpeechLLMs show limited sensitivity to AD-related acoustic-linguistic cues (e.g., abnormal pauses and semantic incoherence) that are critical for diagnosis.
(iv) \textit{Limitations of standard rule-based rewards}: relying solely on rule-based rewards (i.e., format and label correctness) in RL is insufficient to supervise open-ended reasoning quality. This often results in suboptimal reasoning paths and misalignment between the generated reasoning and the predicted label.

To address these challenges, we propose \textbf{ADReasoner}, an RL–enhanced SpeechLLM that generates accurate diagnostic  predictions with interpretable rationales grounded in fine-grained acoustic-linguistic cues (see Figure~\ref{fig:classifier_vs_sft_vs_rl}). 
ADReasoner is developed through four steps. 
First, we fine-tune four LLMs and one Text-to-Speech (TTS) model to generate text-speech pairs for data augmentation. 
We then build an automated annotation pipeline that extracts 12 AD-related acoustic-linguistic cues and uses carefully designed prompts to construct \textbf{ADCoT-7K}, a CoT dataset with diagnostic rationales.
Next, we apply SFT to develop \textbf{ADReasoner-Base}, a cue-perception-enhanced foundation model based on Qwen2.5-Omni-7B \cite{xu2025qwen25omni}, improving AD speech understanding. Finally, we employ Group Relative Policy Optimization (GRPO) for RL. To optimize open-ended reasoning, in addition to the format and label rewards, we introduce a reasoning reward and a Reasoning-Label Consistency (RLC) reward. 
The former evaluates semantic similarity to reference rationales and the coverage of critical cues, while the latter leverages soft probabilities to estimate whether the generated rationale supports the predicted label.
Since open-ended rewards are typically noisier than rule-based ones, we introduce two stabilization mechanisms: a shrinkage mechanism to suppress misleading reasoning signals when incorrect predictions receive high reasoning rewards, and a gating mechanism that progressively 
activates both rewards only after the policy model reaches a predefined detection performance level. In summary, our main contributions are:
\begin{itemize}
    \item We fine-tune four LLMs and one TTS model to generate text-speech pairs, thereby alleviating data scarcity.
    \item We systematically construct ADCoT-7K, an AD speech reasoning dataset that combines real data with our quality-filtered synthetic data, where each instance is enriched with CoT annotations and detailed captions. 
    \item We leverage SFT to develop ADReasoner-Base, a cue-perception-enhanced foundation model that substantially improves the understanding of AD speech.
    \item We introduce reasoning and RLC rewards into GRPO, coupled with two stabilization mechanisms, yielding ADReasoner, the first RL-enhanced SpeechLLM tailored for explainable AD speech reasoning. Experiments on three AD datasets demonstrate that ADReasoner consistently outperforms prior state-of-the-art models in both detection performance and explanation quality.
\end{itemize}

\section{Related Work}
\subsection{Speech-Based AD Detection}
Early speech-based AD detection methods relied mainly on handcrafted acoustic and linguistic features with traditional machine learning classifiers \cite{fraser2016linguistic,de2020artificial,mei2023ustc}. The field later shifted toward fine-tuning pretrained speech models (e.g., Wav2Vec2) for binary classification \cite{liu24f_interspeech, gao2025leveraging}. More recently, researchers have explored SpeechLLMs through zero-shot prompting or SFT \cite{shahin2025zero,liu2026cross}. However, these methods largely remain black boxes, producing binary labels without interpretable rationales. Although a few recent studies \cite{feng2025cogbench,feng2026cross} generate diagnostic labels with explanations via SFT, 
they neither enable autonomous exploration of optimal reasoning paths nor ground their reasoning in fine-grained acoustic and linguistic cues.
Consequently, their rationales remain largely surface-level. These limitations motivate advancing AD detection from classification toward deep, 
explainable reasoning.

\subsection{RL for Speech Reasoning}
Recent advances in RL have shown strong effectiveness in enhancing reasoning across multiple domains \cite{yang2025r1,wang2026emotionthinker}. DeepSeek-R1 \cite{guo2025deepseek} established that RL, even with simple rule-based rewards, can incentivize robust reasoning without the need for dense supervision, providing a scalable paradigm. Building on this, 
R1-AQA \cite{li2025reinforcement} applied RL to speech reasoning and showed advantages over SFT in audio question answering. Subsequent models like Omni-R1 \cite{rouditchenko2025omni} and AudioMCQ \cite{he2025measuring} utilized GRPO to optimize format and answer accuracy. However, these general-purpose methods typically leave intermediate reasoning unconstrained. Accurate AD speech diagnosis heavily depends on the precise integration of specialized acoustic-linguistic cues during the reasoning process. 
Therefore, additional mechanisms are needed to supervise intermediate reasoning and ensure these cues are captured accurately and comprehensively.

\begin{figure*}[t]
    \centering
    \includegraphics[width=1\linewidth]
    {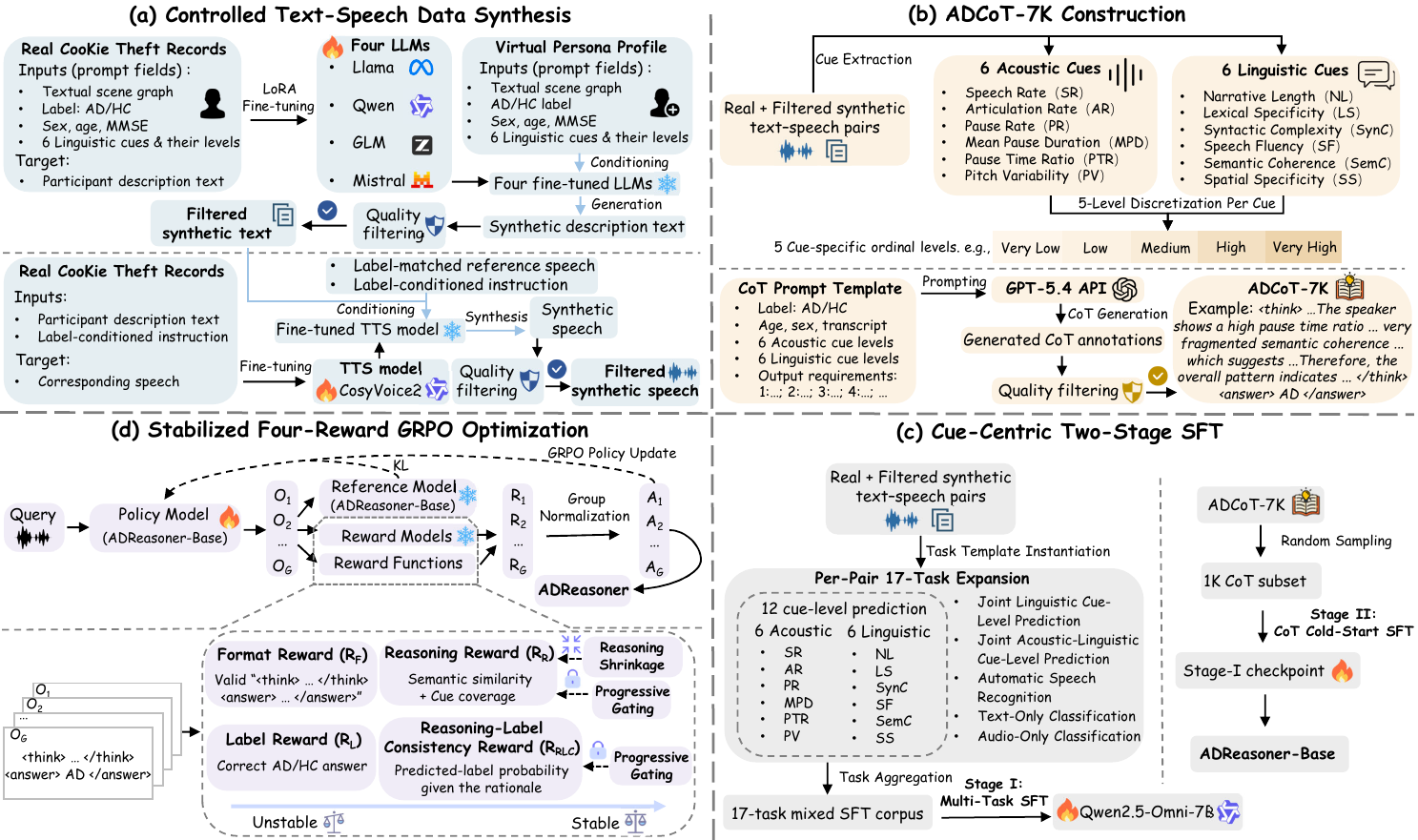}
    \caption{Overview of the four steps in ADReasoner: (a) controlled multi-LLM text generation and CosyVoice2 speech synthesis with text/speech quality filtering; (b) construction of ADCoT-7K by grounding GPT-generated diagnostic reasoning in five-level acoustic and linguistic cues; (c) cue-centric two-stage SFT with a 17-task perception curriculum followed by a CoT cold-start; and (d) GRPO optimization with four rewards, stabilized by reasoning shrinkage and progressive gating.}
    \label{fig:overview}
\end{figure*}

\section{Methods}
ADReasoner formulates explainable AD detection as a generation task. Given an audio recording $a$ from a participant and a task query $q$, the model generates an interpretable rationale $\hat{r}$, followed by a binary diagnosis label $\hat{y}\in\{\mathrm{AD},\mathrm{HC}\}$:
\begin{equation}
o=(\hat{r},\hat{y}) \sim p_{\theta}(\cdot \mid a,q).
\end{equation}
The model output follows a structured format:
\texttt{<think>...\allowbreak</think><answer>AD|HC</answer>}.
As illustrated in Figure~\ref{fig:overview}, ADReasoner comprises four steps: Controlled Text-Speech Data Synthesis, ADCoT-7K Construction, Cue-Centric Two-Stage SFT, and Stabilized Four-Reward GRPO Optimization. We detail each step in the following sections.

\subsection{Controlled Text-Speech Data Synthesis}
AD speech data is highly limited in scale. We therefore construct a synthesis pipeline to enhance training data diversity and alleviate data scarcity (as shown in Figure~\ref{fig:overview}(a)).

\subsubsection{Text Generation}
We fine-tune four LLMs on real Cookie Theft picture descriptions: Llama-3.1-8B-Instruct \cite{grattafiori2024llama}, Qwen3-8B \cite{yang2025qwen3}, GLM-4-9B \cite{glm2024chatglm}, and Mistral-7B-Instruct-v0.3 \cite{Jiang2023Mistral7}. 
Specifically, 
the input prompt includes the picture's textual scene graph, AD/HC label, sex, age, Mini-Mental State Examination (MMSE) score, and six linguistic cues: narrative length, lexical specificity, syntactic complexity, speech fluency, semantic coherence, and spatial specificity, with their ordinal levels, while the description text serves as the target output.

During generation, each fine-tuned LLM produces synthetic descriptions at 10$\times$ the size of the real training set. For the $k$-th model, the synthetic description $t$ is obtained as:
\begin{equation}
t
\sim
p_{\phi_k}
\left(
\cdot
\mid
I,y,d,l
\right),
\end{equation}
where $I$ denotes the textual scene graph, $y$ denotes the target label, $d$ denotes the demographic context (sex, age, MMSE) sampled from same-label real training samples, and $l$ denotes the linguistic-cue profile. Specifically, $l$ is obtained by combining a demographic-conditioned linear model with residuals sampled from real data, and is further mapped into ordinal levels ranging from 1 to 5, which helps preserve within-class heterogeneity. 
The generated texts undergo multi-stage filtering, including format checking, diversity evaluation, and binary-classifier-based filtering. 
The final filtered synthetic texts are then used for speech synthesis.

\subsubsection{Speech Synthesis}
We fine-tune the LLM module of the CosyVoice2 \cite{du2024cosyvoice} TTS model using real text-speech pairs, while keeping the Flow and HiFiGAN modules frozen. 
Specifically, the description text and the designed label-conditioned instruction are used as inputs, and the corresponding speech is used as the output for fine-tuning. During speech synthesis, the synthetic speech $s$ is generated as:
\begin{equation}
s
\sim
p_{\psi}
\left(
\cdot
\mid
t_{\mathrm{tar}},s_{\mathrm{ref}},c
\right),
\end{equation}
where $t_{\mathrm{tar}}$ denotes the target text from the filtered synthetic text corpus, $s_{\mathrm{ref}}$ represents the label-matched reference speech sampled from real speech data, and $c$ denotes the label-conditioned instruction.
The synthesized speech is further processed by multi-stage quality filtering, including duration checking, ASR--text agreement evaluation, and binary-classifier-based filtering. 
Finally, we obtain 6,545 filtered synthetic text-speech pairs, which are used for subsequent CoT construction and cue-centric two-stage SFT. More details about data synthesis are provided in the Appendix B.

\subsection{ADCoT-7K Construction}
Existing AD speech datasets merely contain discrete diagnostic labels, which cannot provide supervision for structured reasoning. To facilitate explainable AD detection, we construct ADCoT-7K, the first richly annotated AD speech reasoning dataset, containing approximately 7,000 speech-reasoning pairs spanning about 110 hours of audio. 

As shown in Figure~\ref{fig:overview}(b), we design an automated annotation pipeline that first extracts six acoustic cues and six linguistic cues from each real and synthetic text-speech pair using standard speech and NLP tools. These cues have proven effective for AD detection \cite{fraser2015linguistic, rohanian21_interspeech}.
The acoustic cues are Speech Rate (SR), Articulation Rate (AR), Pause Rate (PR), Mean Pause Duration (MPD), Pause Time Ratio (PTR), and Pitch Variability (PV); the linguistic cues are Narrative Length (NL), Lexical Specificity (LS), Syntactic Complexity (SynC), Speech Fluency (SF), Semantic Coherence (SemC), and Spatial Specificity (SS). The continuous value of each cue is discretized into five cue-specific ordinal levels based on thresholds estimated from the real training set. Subsequently, the pipeline integrates the extracted cue levels, label, age, sex, transcript, and output requirements into a structured prompt for GPT-5.4 to generate CoT reasoning traces in the format of \texttt{<think>...</think><answer>AD|HC</answer>}. 
The generated CoT traces further undergo quality filtering, including output format checking, length verification, and reasoning-label consistency checking, resulting in the final ADCoT-7K. Additional 
details are provided in Appendix C.

\subsection{Cue-Centric Two-Stage SFT}
\label{sec:two_stage_sft}
Our preliminary experiments (Figure~\ref{fig:sft_stage_task_performance}) show that Qwen2.5-Omni-7B lacks fine-grained perception of AD-related acoustic-linguistic cues. 
However, accurate cue perception is a prerequisite for reliable AD speech reasoning. 
For example, human experts typically assess whether a speech sample may indicate AD by considering multiple cues, such as frequent pauses and semantic incoherence.
Therefore, prior to RL, we introduce cue-centric two-stage SFT to enhance the cue perception of Qwen2.5-Omni-7B (Figure~\ref{fig:overview}(c)).

\subsubsection{Stage I: Multi-Task SFT}
In Stage I, we conduct multi-task SFT with 17 designed tasks. Specifically, these tasks include six acoustic and six linguistic cue-level prediction tasks, where the model learns to assign each cue to one of five ordinal levels directly from the audio; a Joint Linguistic Cue-Level Prediction (JLCLP) task, where the model jointly predicts the levels of six linguistic cues and the diagnostic label from a transcript; a Joint Acoustic-Linguistic Cue-Level Prediction (JALCLP) task, where the model jointly predicts the levels of six acoustic cues, six linguistic cues, and the diagnostic label from audio; an Automatic Speech Recognition (ASR) task, where the model learns to generate transcripts from audio; a Text-Only Classification (TOC) task, where the model predicts the diagnostic label from a transcript; and an Audio-Only Classification (AOC) task, where the model predicts the diagnostic label from audio. For each real and synthetic text-speech pair, we construct 17 task-specific samples, resulting in a 17-task mixed SFT corpus containing approximately 120K samples and spanning about 1,700 hours of audio. We use this mixed SFT corpus to perform full-parameter fine-tuning of the audio encoder, audio adapter, and LLM backbone of Qwen2.5-Omni-7B.

\subsubsection{Stage II: CoT Cold-Start SFT}
In Stage II, we randomly sample a 1K subset from ADCoT-7K and use this subset for cold-start training on the Stage-I checkpoint. 
Specifically, we apply LoRA fine-tuning to the LLM backbone while keeping the audio encoder and audio adapter frozen. 
This enables the model to learn basic \texttt{<think>--<answer>} reasoning patterns. Finally, we obtain the cue-perception-enhanced model ADReasoner-Base.
Additional details about cue-centric two-stage SFT are provided in Appendix D.

\subsection{Stabilized Four-Reward GRPO Optimization}
We leverage GRPO \cite{Shao2024DeepSeekMathPT} to further optimize ADReasoner-Base for improved detection performance and reasoning capability. 
Figure~\ref{fig:overview}(d) illustrates the framework. 
In addition to format and label rewards, we introduce a reasoning reward and an RLC reward into GRPO, together with two stabilization mechanisms, to supervise reasoning quality.

\subsubsection{Format Reward}
The format reward $R_F$ checks whether an output $o$ follows the predefined format, with reasoning in \texttt{<think>...</think>} and the prediction in \texttt{<answer>...</answer>}:
\begin{equation}
R_F(o)=
\begin{cases}
1, & \text{if } o \text{ follows the predefined format},\\
0, & \text{otherwise}.
\end{cases}
\end{equation}

\subsubsection{Label Reward}
The label reward $R_L$ supervises the correctness of the predicted label. 
Given a ground-truth label $y^{\star}$ and a predicted label $\hat{y}$, $R_L$ is defined as:
\begin{equation}
R_L(\hat{y},y^{\star})=
\begin{cases}
1, & \text{if } \hat{y}=y^{\star},\\
0, & \text{otherwise}.
\end{cases}
\end{equation}

\subsubsection{Reasoning Reward}
The reasoning reward $R_R$ encourages evidence-grounded rationale generation by considering rationale similarity and cue coverage. 
Given a generated rationale $\hat{r}$ and a reference rationale $r^{\star}$, the $R_R$ is defined as:
\begin{equation}
R_R(\hat{r})=
\lambda_s S(\hat{r},r^{\star})
+
\lambda_c \mathrm{Cov}(d_{\hat{r}},d_{r^{\star}}),
\end{equation}
where $S(\hat{r},r^{\star})$ denotes the semantic similarity between $\hat{r}$ and $r^{\star}$ computed using cosine similarity over sentence embeddings extracted by a Sentence-BERT model \cite{reimers-gurevych-2019-sentence}.
$d_{r^{\star}}$ denotes the paraphrased cue dictionary derived from acoustic-linguistic cues in $r^{\star}$, and $d_{\hat{r}}$ denotes the cue expressions extracted from $\hat{r}$. 
$\mathrm{Cov}(d_{\hat{r}},d_{r^{\star}})$ denotes the cue coverage ratio, calculated as $|\mathrm{Match}(d_{\hat{r}},d_{r^{\star}})|/|d_{r^{\star}}|$, where a cue is covered if its expression matches a paraphrased expression in $d_{r^{\star}}$. 
Here, $\lambda_s$ and $\lambda_c$ are weighting factors.

\subsubsection{Reasoning-Label Consistency Reward}
Rule-based rewards alone may cause inconsistencies between the generated rationale and the final prediction (Table~\ref{tab:grpo_ablation_design_results_5_seeds_Pitt_Appendix}).
Therefore, we introduce the RLC reward $R_{RLC}$, which encourages the generated rationale to be consistent with the final prediction. 
Unlike the label reward $R_L$, which only evaluates the correctness of the predicted label, $R_{RLC}$ measures whether the rationale supports the predicted diagnosis. 
Given the generated rationale-label pair $(\hat{r},\hat{y})$, $R_{RLC}$ is defined as:
\begin{equation}
R_{RLC}(\hat{r},\hat{y})=P(\hat{y}\mid\hat{r}),
\end{equation}
where $P(\hat{y}\mid\hat{r})$ denotes the probability that rationale $\hat{r}$ supports label $\hat{y}$, estimated by a BERT-based binary classifier trained on rationale-label annotations from ADCoT-7K.

\subsubsection{Reasoning Shrinkage Mechanism}
Although the reasoning reward $R_R$ supervises open-ended reasoning, it is not directly linked to label correctness and may be unreliable when incorrect responses receive higher rewards than correct ones within a GRPO group. 
We therefore introduce a shrinkage factor $\alpha$ to adaptively suppress such misleading signals.
Within each group, let $\bar{R}_{R}^{\mathrm{c}}$ and $\bar{R}_{R}^{\mathrm{inc}}$ denote the mean reasoning rewards of correct and incorrect responses, respectively:
\begin{equation}
\alpha=
\begin{cases}
\exp(\bar{R}_{R}^{\mathrm{c}}-\bar{R}_{R}^{\mathrm{inc}}),
& \text{if } \bar{R}_{R}^{\mathrm{inc}}>
\bar{R}_{R}^{\mathrm{c}},\\
1, & \text{otherwise}.
\end{cases}
\end{equation}
The adjusted reward is $\widetilde{R}_{R}=\alpha R_R$, thereby improving reasoning reward reliability during GRPO optimization.

\subsubsection{Progressive Gating Mechanism}
Reasoning-related rewards are harder to optimize due to the diversity of open-ended generation, and introducing them jointly with rule-based rewards from the start may destabilize reward updates. 
To ensure stable multi-reward optimization, training initially uses only the rule-based rewards $R_F$ and $R_L$. 
Once the policy model reaches a predefined validation accuracy threshold $\delta$, the reasoning rewards $R_R$ and $R_{RLC}$ are gradually activated. 
We define a gating coefficient $g_k(t)$ as:
\begin{equation}
g_k(t)=
\begin{cases}
0, & t<t_{\delta},\\
\min\left(1,\frac{t-t_{\delta}+1}{\gamma}\right), & t\geq t_{\delta},
\end{cases}
\quad
k\in\{\mathrm{R},\mathrm{RLC}\},
\end{equation}
where $t$ is the global training step, $t_{\delta}$ is the first step at which validation accuracy reaches $\delta$, and $\gamma$ is the number of warm-up steps. 
The gating coefficients $g_R$ and $g_{RLC}$ share the same trigger and schedule but can be disabled independently.

\begin{table*}[t]
\centering
\small
\setlength{\tabcolsep}{3.2pt}
\renewcommand{\arraystretch}{0.94}
\begin{threeparttable}
\begin{tabular}{@{}lc*{6}{c}@{\hspace{5pt}}*{6}{c}@{}}
\toprule
\multirow{2}{*}{\textbf{Model}}
& \multirow{2}{*}{\textbf{Year}}
& \multicolumn{6}{c}{\textbf{Pitt Test Set (In-Domain)}}
& \multicolumn{6}{c}{\textbf{PROCESS-1/PROCESS-2 Average (OOD)}} \\
\cmidrule(lr){3-8}\cmidrule(lr){9-14}
& & F1$\uparrow$ & Acc$\uparrow$ & FPR$\uparrow$ & R$\uparrow$ & RLC$\uparrow$ & LLM-R$\uparrow$
& F1$\uparrow$ & Acc$\uparrow$ & FPR$\uparrow$ & R$\uparrow$ & RLC$\uparrow$ & LLM-R$\uparrow$ \\
\midrule

\multicolumn{14}{c}{\textit{Conventional Supervised Classifiers (Non-Reasoning \& Classification)}} \\
\midrule
eGeMAPS + Na\"{\i}ve Bayes & 2023
& 63.85 & 65.69 & - & - & - & -
& 51.44 & 57.37 & - & - & - & - \\

Wav2Vec2 + Linear & 2024
& 73.43 & 74.86 & - & - & - & -
& 59.59 & 67.30 & - & - & - & - \\

Whisper + MLP & 2025
& 72.79 & 73.58 & - & - & - & -
& 59.01 & 64.93 & - & - & - & - \\

Qwen2.5-Omni-7B & 2026
& 74.70 & 75.23 & - & - & - & -
& 60.65 & 66.53 & - & - & - & - \\

\midrule
\multicolumn{14}{c}{\textit{Reasoning-SFT SpeechLLMs (Unstructured Reasoning \& Classification)}} \\
\midrule

Qwen2.5-Omni-7B & 2026
& \underline{75.42} & \underline{75.78} & - & 71.22 & 76.47 & 3.12
& 62.36 & 68.51 & - & 68.96 & 75.53 & 3.02 \\

\midrule
\multicolumn{14}{c}{\textit{Zero-Shot SpeechLLMs (Zero-Shot Reasoning \& Classification)}} \\
\midrule

Qwen-Audio-Chat & 2023
& 44.88 & 46.06 & 35.05 & 34.15 & 45.40 & 1.94
& 45.82 & 60.18 & 34.12 & 36.11 & 62.55 & 2.06 \\

GLM-4-Voice & 2024
& 47.05 & 48.26 & 38.35 & 44.27 & 52.78 & 1.94
& 44.15 & 50.77 & 41.11 & 49.51 & 56.89 & 1.96 \\

Qwen2-Audio-7B-Instruct & 2024
& 37.52 & 42.20 & 32.11 & 46.30 & 50.97 & 1.93
& 52.67 & 56.10 & 31.18 & 43.26 & 60.40 & 1.93 \\

SALMONN & 2024
& 33.48 & 38.17 & 30.09 & 31.74 & 40.32 & 1.50
& 43.39 & 45.18 & 28.18 & 39.45 & 50.89 & 1.57 \\

Megrez-3B-Omni & 2025
& 44.53 & 47.16 & 33.03 & 41.66 & 52.29 & 1.33
& 51.40 & 57.07 & 30.18 & 33.67 & 56.37 & 1.05 \\

MERaLiON-2-10B & 2025
& 47.39 & 59.08 & 63.12 & 57.28 & 55.04 & 2.20
& 56.26 & 60.03 & 60.60 & 57.47 & 59.27 & 2.19 \\

MiMo-Audio-7B-Instruct & 2025
& 57.50 & 64.04 & 75.60 & 64.25 & 73.48 & 2.69
& 63.12 & 67.57 & 73.66 & 63.01 & 69.79 & 2.71 \\

Phi-4-Multimodal-Instruct & 2025
& 25.21 & 31.01 & 26.97 & 33.45 & 39.98 & 1.78
& 34.99 & 35.15 & 23.13 & 34.16 & 45.28 & 2.09 \\

Qwen2.5-Omni-7B & 2025
& 54.39 & 55.05 & 69.72 & 53.84 & 53.08 & 2.67
& 58.87 & 61.55 & 67.61 & 53.35 & 56.69 & 2.76 \\

Qwen3-Omni-30B-A3B-Instruct & 2025
& 57.91 & 67.16 & 81.10 & 68.23 & 69.27 & 2.96
& \underline{64.69} & 68.28 & 82.18 & 64.75 & 73.25 & 2.88 \\

Ultravox-v0.6-Llama-3.1-8B & 2025
& 56.49 & 66.06 & 79.08 & 62.78 & 67.82 & 2.36
& 61.45 & 64.56 & 76.40 & 60.30 & 70.32 & 2.29 \\

Voxtral-Mini-3B   & 2025
& 58.72 & 62.20 & 77.06 & 56.34 & 64.23 & 2.25
& 59.88 & 66.06 & 75.05 & 61.58 & 59.46 & 2.33 \\

Audio-Flamingo-Next-Think & 2026
& 44.07 & 49.17 & 45.14 & 51.76 & 56.58 & 1.79
& 47.24 & 55.74 & 52.22 & 56.20 & 59.86 & 1.82 \\

MiniCPM-o-4.5 & 2026
& 61.27 & 68.07 & 79.82 & 66.45 & 70.31 & 3.06
& 63.60 & 67.07 & 78.24 & 67.67 & 72.17 & 2.99 \\

MOSS-Audio-8B-Thinking & 2026
& 34.78 & 43.12 & 42.02 & 54.22 & 58.23 & 2.25
& 49.99 & 51.19 & 50.85 & 56.72 & 59.18 & 2.20 \\

Gemini-3.5-Flash$^{\dagger}$ & 2026
& 59.88 & 63.12 & 85.50 & 73.56 & \underline{82.14} & 3.18
& 53.22 & 73.24 & 86.12 & 75.22 & \underline{82.24} & 3.20 \\

GPT-audio-1.5$^{\dagger}$ & 2026
& 66.32 & 66.42 & \underline{93.76} & \underline{76.03} & 81.25 & \underline{3.27}
& 50.04 & \textbf{75.97} & \underline{89.29} & \underline{75.94} & 81.82 & \underline{3.31} \\

\midrule
\midrule

\textbf{ADReasoner (Ours)} & 2026
& \textbf{80.72} & \textbf{81.10} & \textbf{100.00} & \textbf{85.24} & \textbf{90.86} & \textbf{3.70}
& \textbf{65.65} & \underline{74.38} & \textbf{100.00} & \textbf{81.63} & \textbf{86.03} & \textbf{3.54} \\

\bottomrule
\end{tabular}

\caption{
Performance comparison across models.
F1, Acc, FPR, R, and RLC denote macro-F1, accuracy, format parsing rate, reasoning score, and reasoning-label consistency score, respectively (\%).
LLM-R is the mean 1--5 score over DFA, CIV, CCC, and RSC.
$^{\dagger}$ denotes closed-source models.
Best results are in \textbf{bold}, second-best are \protect\underline{underlined}. 
``-'' indicates not applicable.}
\label{tab:main_pitt_ood_results}
\end{threeparttable}
\end{table*}

\begin{table}[t]
\centering
\small
\setlength{\tabcolsep}{3pt}
\begin{threeparttable}
\begin{tabular}{@{}p{0.44\columnwidth}cc@{}}
\toprule
\textbf{Model} & \textbf{Pitt Test Set} & \textbf{OOD Average} \\
\midrule
Phi-4-Multimodal-Instruct & 2.40 & 2.89 \\
Qwen-Audio-Chat & 2.45 & 2.84 \\
MOSS-Audio-8B-Thinking & 3.37 & 3.67 \\
Qwen2.5-Omni-7B & 3.62 & 3.74 \\
MiniCPM-o-4.5 & 3.67 & 3.45 \\
Gemini-3.5-Flash & 3.86 & 3.81 \\
GPT-audio-1.5 & 3.94 & 3.91 \\
\midrule
\textbf{ADReasoner (Ours)} & \textbf{4.15} & \textbf{4.02} \\
\bottomrule
\end{tabular}%

\caption{Human evaluation results averaged over four dimensions: DFA, CIV, CCC, and RSC.}
\label{tab:human_eval_pitt_ood}
\end{threeparttable}
\end{table}

\subsubsection{Overall Reward}
The overall reward for GRPO optimization is formulated as:
\begin{equation}
R=
w_F R_F+
w_L R_L+
w_R g_R \alpha R_R+
w_{RLC} g_{RLC}R_{RLC},
\end{equation}
where $w_F$, $w_L$, $w_R$, and $w_{RLC}$ denote the weights of the corresponding reward components. $g_R$ and $g_{RLC}$ represent the gating coefficients, and $\alpha$ denotes the shrinkage factor.

\section{Experiments}
\subsection{Experimental Setup}
\subsubsection{Datasets}
We conducted experiments on three widely used AD speech datasets: Pitt \cite{becker1994natural}, PROCESS-1 \cite{tao2025early}, and PROCESS-2 \cite{Pahar2026PROCESS2AB}. 
Pitt was used as the in-domain benchmark, while PROCESS-1 and PROCESS-2 were used only for out-of-domain (OOD) evaluation. 
Pitt contains 552 samples from the Cookie Theft picture description task \cite{goodglass1983boston}, including 309 AD and 243 HC samples. 
We split Pitt into training, validation, and test sets with a participant-independent stratified split ratio of 7:1:2, resulting in 394 (222 AD, 172 HC), 49 (25 AD, 24 HC), and 109 (62 AD, 47 HC) samples, respectively. 
PROCESS-1 and PROCESS-2 contain 98 (16 AD, 82 HC) and 250 (50 AD, 200 HC) samples, respectively. 
More details about datasets are provided in Appendix F.

\subsubsection{Baselines}
We compared ADReasoner with three categories of baselines. 
\textbf{(a) Conventional Supervised Classifiers.}
These methods only output binary labels without reasoning, including eGeMAPS + Na\"ive Bayes \cite{luz2023multilingual}, Wav2Vec2 + Linear \cite{liu24f_interspeech}, Whisper + MLP \cite{gao2025leveraging}, and fine-tuned Qwen2.5-Omni-7B \cite{liu2026cross}. 
\textbf{(b) Reasoning-SFT SpeechLLMs.}
Following the only existing work in this category \cite{feng2026cross}, we reproduced its approach by fine-tuning Qwen2.5-Omni-7B using reasoning data without a predefined XML-style format.
\textbf{(c) Zero-Shot SpeechLLMs.}
We evaluated 15 open-source and two closed-source SpeechLLMs in a zero-shot setting.
The open-source models included 
Qwen-Audio-Chat \cite{chu2023qwen}, 
GLM-4-Voice \cite{zeng2024glm},
Qwen2-Audio-7B-Instruct \cite{chu2024qwen2}, 
SALMONN \cite{tang2024salmonn},
Megrez-3B-Omni \cite{li2025megrez}, 
MERaLiON-2-10B \cite{meralionteam2025meralion2},
MiMo-Audio-7B-Instruct \cite{zhang2025mimo},
Phi-4-Multimodal-Instruct \cite{abouelenin2025phi}, 
Qwen2.5-Omni-7B,
Qwen3-Omni-30B-A3B-Instruct \cite{xu2025qwen3}, 
Ultravox-v0.6-Llama-3.1-8B \cite{fixieai2025ultravoxv06}, 
Voxtral-Mini-3B \cite{liu2025voxtral},
Audio-Flamingo-Next-Think \cite{ghosh2026audio},
MiniCPM-o-4.5 \cite{cui2026minicpm}, 
and MOSS-Audio-8B-Thinking \cite{yang2026moss}.
The closed-source models included
Gemini-3.5-Flash and GPT-audio-1.5.
More details are provided in Appendix G.

\begin{table*}[t]
\centering
\small %
\setlength{\tabcolsep}{1.8pt} %
\begin{tabular}{ccccccccc|cccccc}
\toprule
\multirow{2}{*}{Variant}
& \multirow{2}{*}{Method}
& \multicolumn{7}{c|}{Reward Configuration}
& \multicolumn{6}{c}{Evaluation Metric} \\
\cmidrule(lr){3-9} \cmidrule(lr){10-15}
&
& F
& L
& R
& Shrinkage
& R-gate
& RLC
& RLC-gate
& F1 $\uparrow$
& Acc $\uparrow$
& FPR $\uparrow$
& R $\uparrow$
& RLC $\uparrow$
& LLM-R $\uparrow$ \\
\midrule

Baseline 1
& Qwen2.5-Omni-7B
& - & - & - & - & - & - & -
& 54.39
& 55.05
& 69.72
& 53.84
& 53.08
& 2.67 \\

Baseline 2
& SFT (ADReasoner-Base)
& - & - & - & - & - & - & -
& 75.82
& 76.15
& \underline{96.33}
& 78.35
& 83.31
& 3.30 \\

\midrule

V1
& GRPO-F-L
& \cmark & \cmark & \xmark & \xmark & \xmark & \xmark & \xmark
& 79.20
& 79.45
& \textbf{100.00}
& 80.12
& 85.36
& 3.40 \\

V2
& GRPO-F-L-R-w/o-shrinkage
& \cmark & \cmark & \cmark & \xmark & \cmark & \xmark & \xmark
& 79.38
& 79.82
& \textbf{100.00}
& 83.03
& 86.31
& 3.52 \\

V3
& GRPO-F-L-R-w/o-gate
& \cmark & \cmark & \cmark & \cmark & \xmark & \xmark & \xmark
& 77.53
& 77.98
& \textbf{100.00}
& 82.69
& 85.95
& 3.47 \\

V4
& GRPO-F-L-R
& \cmark & \cmark & \cmark & \cmark & \cmark & \xmark & \xmark
& \underline{80.28}
& \underline{80.55}
& \textbf{100.00}
& 84.83
& 86.12
& 3.65 \\

V5
& GRPO-F-L-R-RLC-w/o-RLC-gate
& \cmark & \cmark & \cmark & \cmark & \cmark & \cmark & \xmark
& 79.95
& 80.37
& \textbf{100.00}
& \underline{85.07}
& \underline{90.43}
& \underline{3.68} \\

V6
& GRPO-F-L-R-RLC (ADReasoner)
& \cmark & \cmark & \cmark & \cmark & \cmark & \cmark & \cmark
& \textbf{80.72}
& \textbf{81.10}
& \textbf{100.00}
& \textbf{85.24}
& \textbf{90.86}
& \textbf{3.70} \\

\bottomrule
\end{tabular}%
\caption{Ablation study on the Pitt test set. V1--V6 are built on Baseline 2. F, L, R, and RLC denote format, label, reasoning, and reasoning-label consistency rewards, respectively. Best and second-best results are in \textbf{bold} and \protect\underline{underlined}.}
\label{tab:grpo_ablation_design_results_5_seeds_Pitt_Appendix}
\end{table*}

\subsubsection{Evaluation Metrics}
We used macro-F1 score (F1) and label accuracy (Acc) to evaluate AD detection performance. 
To assess reasoning quality, we employed format parsing rate (FPR), reasoning score (R), reasoning-label consistency score (RLC), and LLM-based reasoning score (LLM-R). 
Specifically, we used GPT-5.4 to evaluate the generated reasoning outputs from four dimensions: Diagnostic Factual Alignment (DFA), Clinical Interpretative Validity (CIV), Cognitive Cue Completeness (CCC), and Reasoning Structure and Clarity (RSC), assigning each dimension a score from 1 to 5. 
LLM-R is the arithmetic mean of these four scores. 
More details are provided in Appendix H.
We further conducted human evaluation, as described in Section~\ref{sec:human_evaluation}.

\subsubsection{Implementation Details}
ADReasoner-Base was initialized from Qwen2.5-Omni-7B and trained for two epochs in each SFT stage; ADReasoner was then initialized from ADReasoner-Base. The RL post-training was implemented with MS-SWIFT \cite{zhao2025swift}, using vLLM \cite{kwon2023efficient} for accelerated rollout generation. It was conducted on 6K samples from ADCoT-7K for 4,000 steps using four A800 GPUs (80G). 
We used a KL coefficient of 0.04, a learning rate of $2\times10^{-6}$, and eight responses per query. 
The reward weights $(w_F,w_L,w_R,w_{RLC})$ were $(0.10,0.50,0.25,0.15)$, with $\lambda_s=\lambda_c=0.5$, $\delta=0.78$, and $\gamma=200$. Only synthetic data and designated training splits were used for training; all test splits were reserved for evaluation. All experimental results were averaged over five independent runs. 
Additional details are provided in Appendix I.

\begin{figure}[t]
    \centering
    \includegraphics[width=1\linewidth]{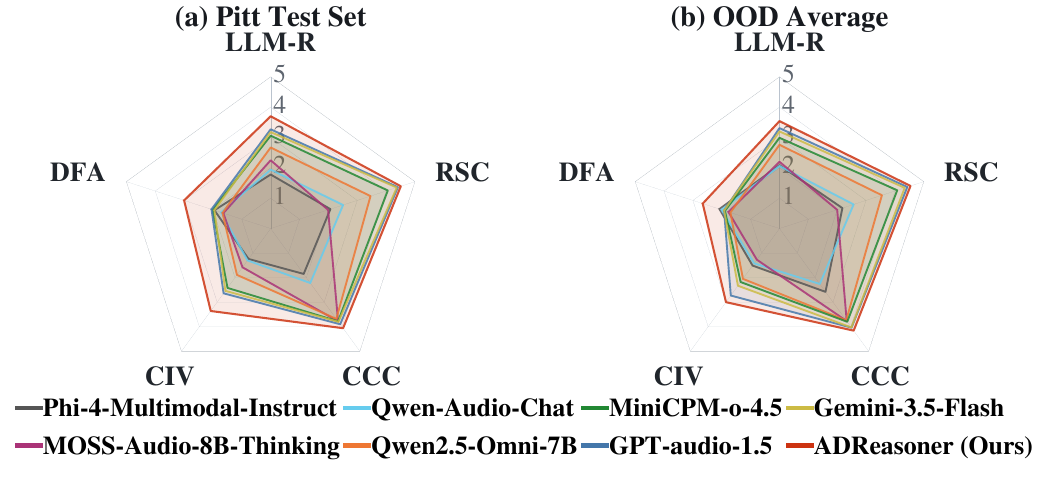}
    \caption{Comparison of LLM-based reasoning evaluation across four dimensions and their average (LLM-R).}
    \label{fig:dual_radar_8_models_pitt_ood}
\end{figure}

\subsection{Results and Analysis}
\subsubsection{Comparison with Prior Baselines}
As shown in Table~\ref{tab:main_pitt_ood_results}, ADReasoner achieves the best performance on both AD detection and reasoning quality on the Pitt test set. For AD detection, ADReasoner obtains 80.72\% F1 and 81.10\% Acc, surpassing the second-best model by approximately 5.3 percentage points on both metrics. For reasoning quality, ADReasoner consistently outperforms all baselines across different metrics. For example, it achieves an LLM-R score of 3.70, yielding a 13.15\% relative improvement over GPT-audio-1.5 (3.27), the strongest baseline.
Notably, most existing baselines lack native structured reasoning, whereas ADReasoner achieves a 100\% FPR while maintaining the highest reasoning quality. 
The OOD results further demonstrate the generalization ability of ADReasoner. 
It achieves the best F1 and all reasoning-related metrics. Although its Acc (74.38\%) is slightly lower than GPT-audio-1.5 (75.97\%), ADReasoner achieves a substantially higher F1 (65.65\% vs. 50.04\%), indicating more balanced diagnostic predictions. 
Figure~\ref{fig:dual_radar_8_models_pitt_ood} compares the four dimensions of LLM-based reasoning evaluation and their average (LLM-R) among eight representative models ranked by their Pitt LLM-R scores. ADReasoner achieves the highest scores across all four dimensions and the highest LLM-R, with particularly stronger performance in Diagnostic Factual Alignment (DFA) and Clinical Interpretative Validity (CIV), demonstrating its superior reasoning quality.
Overall, ADReasoner delivers the strongest comprehensive performance on both in-domain and OOD evaluations, effectively combining accurate AD detection with reliable and interpretable reasoning. Reasoning examples for each model are provided in Appendix K.

\subsubsection{Human Evaluation}
\label{sec:human_evaluation}
To mitigate potential bias in the LLM-based reasoning evaluation, we conducted a human evaluation of reasoning quality. 
Specifically, we randomly sampled 20 class-balanced instances from each of the three datasets. The reasoning outputs of eight representative models on these samples were anonymized and evaluated by five independent reviewers using the same four-dimensional rubric as the LLM-based evaluation. 
As shown in Table~\ref{tab:human_eval_pitt_ood}, ADReasoner achieves the highest human evaluation scores, providing further evidence of its stronger reasoning quality. Moreover, the model ranking closely aligns with the LLM-based evaluation results, indicating robust agreement between human and automatic judgments.

\subsubsection{Cue Perception Comparison and Effectiveness of Synthetic Data}
\begin{figure}[t]
    \centering
    \includegraphics[width=1\linewidth]
    {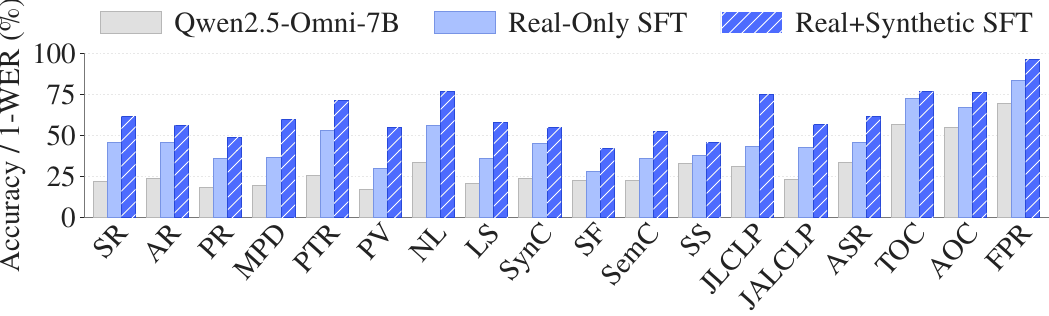}
    \caption{Cue perception comparison. Results are reported as accuracy, except ASR, which uses one minus word error rate (WER). FPR denotes format parsing rate; other task abbreviations follow the definitions in Section~\ref{sec:two_stage_sft}.}
    \label{fig:sft_stage_task_performance}
\end{figure}

We evaluated cue perception on the Pitt test set. 
As shown in Figure~\ref{fig:sft_stage_task_performance}, the original Qwen2.5-Omni-7B exhibits limited perception of AD-related acoustic and linguistic cues. 
Applying cue-centric SFT with real data substantially improves performance across all tasks, confirming the necessity and effectiveness of cue-centric SFT for strengthening cue perception before RL. 
Further incorporating quality-filtered synthetic data (ADReasoner-Base) consistently yields the best results, demonstrating its value in enriching supervision and strengthening AD speech understanding. 
Overall, these findings validate both the proposed cue-centric SFT and the controlled synthetic data pipeline.

\subsubsection{Ablation Study}
We conducted ablation studies on the key components of ADReasoner.
As shown in Table~\ref{tab:grpo_ablation_design_results_5_seeds_Pitt_Appendix}, ADReasoner-Base substantially outperforms the original Qwen2.5-Omni-7B, establishing a strong cue-aware foundation for subsequent RL optimization. 
Using only the format and label rewards (V1) further improves output validity and detection performance, but yields relatively limited gains in reasoning quality, indicating that outcome-level supervision alone is insufficient. 
Introducing the reasoning reward improves all reasoning-related metrics. 
Compared with V2, reasoning shrinkage in V4 increases F1 from 79.38\% to 80.28\% and LLM-R from 3.52 to 3.65, while progressive reasoning gating improves F1 by 2.75 percentage points and LLM-R by 0.18 over V3. 
Adding the RLC reward in V5 further raises the RLC score from 86.12\% to 90.43\%, and progressively activating it in V6 yields additional gains across detection and reasoning metrics. 
Overall, the complete ADReasoner performs best, validating the effectiveness of the proposed reward design, reasoning shrinkage, and progressive gating in improving both AD detection and reasoning quality.

\section{Conclusion}
In this work, we propose ADReasoner, which, to our knowledge, is the first RL-enhanced SpeechLLM to reframe speech-based AD detection from a classification task into interpretable, cue-grounded reasoning.
To alleviate data scarcity, we developed a controlled text-speech synthesis pipeline and constructed ADCoT-7K by integrating real and quality-filtered synthetic data with fine-grained acoustic-linguistic cues and diagnostic reasoning annotations. 
We further introduced cue-centric SFT to build ADReasoner-Base with enhanced cue perception. 
Building on this base model, our stabilized four-reward GRPO framework directly optimizes reasoning quality and reasoning-label consistency through dedicated rewards, reasoning shrinkage, and progressive reward gating. 
Experiments on three AD speech datasets demonstrate that ADReasoner achieves state-of-the-art performance in both AD detection and reasoning quality.

\section*{Acknowledgments}
This work is supported in part by the National Social Science Foundation of China (Grant No. 23AYY012) and by the Supercomputing Center of the USTC.

\bibliography{ADReasoner_references}

\end{document}